\documentclass[12pt]{article}
\usepackage{amsthm}
\usepackage{amsmath,amssymb,enumerate} 
\usepackage{program}
\newtheoremstyle{bthm}{\baselineskip}{\baselineskip}{\slshape}{}{\bfseries}{:}{\newline }{}
\newtheoremstyle{bex}{\baselineskip}{\baselineskip}{}{}{\sffamily}{:}{\newline }{}
\newtheoremstyle{bproof}{\baselineskip}{\baselineskip}{}{}{\scshape}{:}{\newline }{}
\theoremstyle{bthm}

\theoremstyle{bex}

\theoremstyle{bproof}

\newpage
\title{\textbf{\small An introduction to functional dependency in relational databases \\
}}  
\bigskip  \bigskip  \bigskip \bigskip  \bigskip  \bigskip
\author{\small
K.V.Iyer \\  
\small Computer Science and Engineering \\  
\small National Institute of Technology \\
\small Tiruchirapalli -- 620015\\  \\  \\  \\  \\ \\ \\   \\  \\   \\   \\   \\  
 \\   \\   \\  \\  \\  \\  \\  \\
} 
\date{\small  \textsf{Review of relational databases \\
February 2016}}
\newpage  
\begin{document}
\thispagestyle{empty}
\maketitle
\newpage
\vspace*{30mm}
\tableofcontents
\newpage
\vspace*{3cm}
\vspace*{10mm}
\begin{quote} \begin{quote}
\begin{enumerate}
\item[\textsc{Note:}]
These are parts of suggested lecture notes for a second level course on advanced topics in database systems suitable for master's students of Computer Science with a theoretical background. A prerequisite in algorithms and an exposure to data-base systems are required. Additional reading may require exposure to mathematical logic. 
The starting source of this write-up is a survey by M.Y.Vardi listed as reference \cite{vard} - some of the proofs are presented as in \cite{vard}.
This write-up may be considered as a beginning point to eventually lead to other topics in database dependency theory more appropriate for researchers. Database dependency theory is a very rich field that has  bearing on more practical concerns pertaining to implementations. As a consequence modern researchers have considered deviations from classical approaches to suggest fixed-parameter algorithms under restricted cases for certain intractable design problems.     
 \end{enumerate} 
\end{quote}  \end{quote}
\maketitle
\newpage
\setcounter{page}{1}
\pagenumbering{arabic}
\vspace*{10mm}
\section{Introduction}
Manipulation of a large store of structured information has been the fundamental requirement in many computer-based applications which has evolved into database systems and has promoted the associated technologies in the West.  A database management system is now understood to be a computer-based system  maintaining a large amount of permanent data appertaining to a real-world organization/institution together with mechanisms to search, add, update, delete data and with mechanisms for administrative control such as granting/revoking privileges, defining views for restricted access and archieving. In the business domain relational database story has been a success due to cost-effectiveness and the support of a sound formalism in organizing and managing structured data. IBM's pioneering implementation during the 1970s of these concepts was System R.  As apparent from the contemporary literature relational databases still form the core in most of the database management systems since their introduction more than three decades before.\\*
Conceived in the late 1960's, the relational model of databases views a database as a collection of relations where each relation is a set of well-defined tuples. A relation is synomymous with a table whose columns are named by attributes; the rows or tuples capture the real-world information. Since the available information in general is incomplete, \textit{null values} may be permitted in tuples. This notion of databases apparent from the work of E.F.Codd in 1970's is founded on the following two principles: 
\begin{enumerate}[{(a)}]
\item All information pertaining to an application are captured as data values in relations or tables.
\item No information is represented by ordering of columns or rows of any table.
\end{enumerate}
Searching, adding, deleting and updating of data are effected by manipulations of relations by relational algebra having a procedural flavor or by relational calculus having a declarative flavor. Codd's theorem states that any relational algebra expression can be converted efficiently to an equivalent relational calculus expression and \textit{vice versa}. Here efficiency is interpreted to mean that there exists a conversion algorithm whose running time is bounded by a polynomial in the size of the input expression. The relational model is almost devoid of semantics. Therefore meaningful relations in a given context are understood by specifying semantic or integrity constraints. In particular the notion of \textit{functional dependency} introduced by Codd in 1972 is of significance in practice in the sense that a database at no time can misrepresent real-world affairs. The notion of functional dependency is general to information systems in the sense that it applies to a large class of data management tasks (see \cite{pareda, sven1}). The related concept of \textit{implication} apparent from the work of P.A.Bernstein in now considered fundamental in unifying many concepts relevant to database design. The goal in relational database schema design is to formally capture the features of an underlying application at the design level together with concerns for optimization. As pointed out in \cite{jobis2} a schema design theory is to investigate the associated guiding principles and also ``to provide insight into how desirable syntactic properties of schemas are related to worthwhile semantic properties, how desirable syntactic properties can be decided or achieved algorithmically or how the syntactic properties determine costs of storage, queries and updates". Sight should not be lost of the fact that fixed at the design stage, a schema is time-independent together with time-varying instances describing the structure of the data and possibly the associated semantics related to the underlying application. Contributing to the
design formalism, subsequently in 1976 multivalued dependency was introduced by R.Fagin and C.Zaniolo independently, opening up the way for data dependency analyses.  Much of the theory about the relational model during the 1970's and first half of 1980's focussed on query processing and optimization and database design including dependency analysis (see \cite{yanna}). Another studied problem relating to integrity constraints that allow permissible data in relation instances is this: assuming that the currrent data values satisfy the constraints before an update, how to efficiently check (or decide that no checking is necessary) that the constraints hold after the update? \\*
It is known that functional dependencies are fundamental to relational database modeling and design upto BCNF. Experts opine that the theory of functional dependency can be reused in other contexts e.g., in {\it Extended Entity-Relationship} models. This condensed survey is a select rewrite stemming from M.Y.Vardi's survey  \cite{vard} incorporating more explanations as needed from other cited references. Further elaborations as regards issues in database design and other pertinent technical details, associated concepts such as other types of dependencies and their significance in the real-world, intractability results, relationships to mathematical logic and combinatorial problems such as constraint satisfaction and many original references can be found in the database literature from the 1970's (see for example \cite{jobis2} with $103$ listed references) and more specialised ones such as \cite{fag, hmani, btoll, vard, yanna}).
\section{Preliminaries}
In the sequel, it is understood that the implicit context is a given real-world application. We use $I, J, K, \ldots$ to denote the different tables that together form a database. The set $X$ of all attributes in a relation $I$ is referred to as a relation scheme. We then say $I$ is defined over $X$.  By convention we denote by $U$ the set of all attributes occuring across all tables comprising the database. The headers $A,B,C, \ldots$ denote attributes. The tailenders $R,S, \ldots ,X,Y,Z$ denote sets of attributes. For the sake of convenience we make no distinction between $\{ A \}$ and $A.$  Given the attribute sets $X$ and $Y$, $XY$ will denote $X \cup Y$; $ACE$ is a shorthand for $\{A, C, E \}$. Associated with each column of a table is a domain of values from which the entries are taken in the column. That is, for each attribute we have a finite or infinite set $Dom(A)$. In the relational model the elements of $Dom(A)$ are assumed to be atomic in the usual sense and attributes $A,B$ are said to be {\it compatible} if $Dom(A)=Dom(B)$. We denote \textsc{Dom} = $\bigcup_j Dom(A_j)$ where $U= \{A_1, \ldots , A_n \}$. We recall that by tuple we mean a row in a table, that specifies an appropriate value for each attribute. By $u,v,w, \ldots$ we denote tuples. Given an attribute set $X$, a tuple $u$ on $X$ is then a mapping $u: \ X \rightarrow $ \textsc{Dom} such that for each $A$, $u(A) \in Dom(A)$. When incomplete information is available we let some attributes in $X$ take the {\it null value,} denoted by $\O$. A tuple $u$ on $X$ is said to be $A$-$total$ if $u(A) \ne \O$; it is $X$-$total$ if it is $A$-$total$ for every $A$ in $X$.  $\vert X \vert$ can be as large as the \emph{arity} or the total number of attributes in the relation under consideration. If $v$ is a tuple on $X$ then $v[Y]$ will mean the restriction of $v$ to $Y$ where $Y \subset X$. We take $v[X]=v$. \\*
Ex. We can consider relations defined over $ACE$ with $Dom(\alpha)=\{0,1\}$ where $\alpha=A,C$ or $E$.
\subsection{Projection and join} 
Let $X$ be a relation scheme and $I$ a relation on $X$. Given 
$Y \subset X$ we define the \textit{projection} $\pi_Y(I)$, a relation as 
              $\pi_Y(I)=\{ w[Y]\, \vert \, w \in I \}$. 
Let $I_1, \ldots , I_k$ be relations on $X_1, \ldots ,X_k$ and let $X = \cup_{j=1}^{k}X_j$.
Then we define the \textit{join} $I_1 \Join \ldots \Join I_k$ abbreviated to $\Join_{j=1}^{k} I_j$ as 
$\Join_{j=1}^{k} I_j = \{ w[X]\, \vert \, w[X_j] \in I_j \; \forall j \; 1 \leq j \leq k \}$.

Projection builds a new relation from a given one by selecting one
or more attributes. Join combines tuples from two or more relations when 
they agree on common columns. Join is commutative as well as associative.
In some sense projection and join are duals. Beginning with the following relations the examples below illustrate that projection and join cannot always be regarded as inverses. 
{\small
\begin{verbatim} 
      I:  -------   J:  -------    K: ----   L:  ----    M: -------   N: ---- 
          A  B  C       A  B  C       A  B       B  C       A  B  C      A  B 
          -------       -------       ----       ----       -------      ----          
          0  0  0       0  0  0       0  0       0  0       0  0  0      0  0 
          1  0  1       1  0  1       0  1       0  1       0  0  1      ----
          0  0  1       -------       ----       ----       ------- 
          1  0  0  
          ------- 
\end{verbatim}
}   
\noindent
We have $\pi_{AB}(I) \Join \pi_{BC}(I)=I$ and  $\pi_{AB}(J) \Join \pi_{BC}(J)=I$. Also $K \Join L=M$ while $\pi_{AB}(M)=N$ and $\pi_{BC}(M)=L$. 

\noindent
Generalization to the following lemma is immediate.  \\
\textbf{Lemma 1:}  \\ 
(a) Let $I$  be a relation on $X$. Let $X_1, \ldots , X_m$ be
attribute sets such that $X= \cup_{j=1}^{m}X_j$. Then $I \subseteq \; \Join_{j=1}^{m}\Pi_{X_j}(I)$.  \\
The simultaneous projection of $I$ onto $X_1, \ldots , X_m$ is referred to as a 
\textit{decomposition}. The decomposition is \textit{lossless} when 
$I = \; \Join_{j=1}^{m}\Pi_{X_j}(I)$; otherwise it is \textit{lossy}.  \\
(b) Let $I_1, \ldots , I_m$ be relations on $X_1, \ldots ,X_m$ respectively. Then 
$\Pi_{X_j} \big( \Join_{k=1}^{m}I_k \big) \subseteq I_j$. \\*
\textbf{Remark 1:} It is appropriate to interpret lemma 1 for legal relations i.e., when the relations are meaningful with respect to a given real-world scenario. More details follow. 

\section{Functional dependency}
The presence of redundancy and anomalies in instances of relations and the natural requirement to do away with them has motivated the dependency theory of relational databases and hence database design. We first recall what are commonly referred to as Codd's anomalies in a relation by considering the following example.
\begin{quote}
{\small
\begin{verbatim}
               -------------------------------------                    
               STUDENT     DEPARTMENT     SUPERVISOR
               -------------------------------------
                Alice      Cryptology      John
                Bob        Cryptology      John
                Carol      Graph Theory    Yohann
                Darrel     Graph Theory    Yohann 
                Engels     Cryptology      John
                Frank      Graph Theory    Yohann
                Guthrie    Graph Theory    Yohann
               -------------------------------------     
\end{verbatim}
}
\end{quote}
\noindent
The cited problems are:
\begin{enumerate}[(a)]
 \item  \textit{Redundancy:} That John is a supervisor for
Cryptology or Yohann is a supervisor for graph theory can get repeated in many  tuples. 
 \item  \textit{Potential inconsistency:} In the graph theory department if Carol get a new supervisor does it mean that the department gets two supervisors or is the intended meaning to change the supervisor to the new supervisor for all students in graph theory department ?
\end{enumerate}  
It stands to reason that there is a \emph{functional dependency} between \textsf{DEPARTMENT} and \textsf{MANAGER} -- this is a kind of semantic constraint on the data that comprise legal relations.  In this case we say that  \textsf{DEPARTMENT} determines \textsf{SUPERVISOR} and write \textsf{DEPARTMENT} $\longrightarrow$  \textsf{SUPERVISOR}. \\
In formal terms, for attribute sets $X,Y$ $X \longrightarrow Y$ is a functional dependency (FD) over a relation scheme $R$ i.e., $XY \subseteq R$ if for all tuples $u,v \in I$ $u[X]=v[X] \implies u[Y]=v[Y]$, where $I$ is any legal null-free relation on $R.$ We say $I$ satisfies a set of FD's $\Sigma$ if $I$ satisfies all FD's in $\Sigma.$. Specification of an FD $X \longrightarrow Y$ imples that any FD $X \longrightarrow W$ trivially holds whenever $Y \supset W$. A simple FD is one of the form $A \longrightarrow B$. It appears that simple FDs are of interest in the context of data warehouses as pointed out in \cite{jens}. \\
Two notions namely equivalence and redundancy of FDs are useful in the context of manipulating FDs mechanically. Two sets of FDs $\Delta$ and $\Sigma$ are
\textit{equivalent} written $\Sigma \equiv \Delta$ if they are precisely satisfied by the same set of legal relations. In other words $I$ satisfies $\Delta$ $\Longleftrightarrow$ $I$ satisfies $\Sigma$. A set $\Sigma$ of fd's is \textit{redundant} if there is a $\Delta \subset \Sigma$ such that $\Delta \equiv \Sigma$. These can be expressed by a more fundamental notion viz., \textit{implication}. We write $\Sigma \models \sigma$ to say that a set $\Sigma$ of FDs implies an FD $\sigma$. By this we mean that any relation that satisfies $\Sigma$ necessarily satisfies $\sigma$. For example $ \{B \longrightarrow C, A \longrightarrow B \} \models A  \longrightarrow C $.  The database implication (or inference) problem for FDs is: given $\Sigma$ that necessarily holds for any legal null-free instance of a database and given any $\sigma$ does  $\Sigma \models \sigma$?  \\
In terms of implication redundancy and equivalence can be stated as
\begin{enumerate}[(i)]
\item $\Sigma$ is redundant iff there is an FD $\sigma \in \Sigma$ such that $\Sigma -\{\sigma\}$ $\models \sigma$.
\item $\Delta \equiv \Sigma$ iff $\Delta \models \sigma$ for any
$\sigma \in \Sigma$ and $\Sigma \models \delta$ for any $\delta \in \Delta.$
\end{enumerate}
\textbf{Remark 2:}\\
i) We note that $\Sigma$ can be referred to as {\it a set of independent FDs} if it has no redundancy.
ii) If $X \cap Y=\phi$ then the FD $X \longrightarrow Y$
may be interpreted as a case of \textit{multivalued dependency} (see \cite{fag} for example).\\ 
iii) Let $I$ be any relation on $R$. Then an FD $X \longrightarrow Y$ where $XY \subset R$  is satisfied by $I$ iff $\pi_{XY}(I)$ satisfies $X \longrightarrow Y$.  \\
iv) An {\it FD with nulls}, over a relation scheme $R$, $X \longrightarrow Y$ holds in $I$ if for each pair $u,v$ of X-total tuples, $u[X]=v[X] \implies u[Y]=v[Y]$.   \\
v) It has been shown that  FDs can be interpreted as formulas in propositional calculus. To this end it is sufficient to interpret  an FD like $A_1 \ldots A_k \longrightarrow B$ as an equivalent \textit{Horn formula} and then take note of the fact that for Horn formulas satisfiability can be tested in polynomial-time. In turn this implies the existence of a polynomial-time algorithm for the implication problem.

In the above finite as well as infinite relations are allowed though in practice we need to consider only finite relations. Written as $\Sigma \models_f \sigma$, $\Sigma$ finitely implies $\sigma$ if any finite relation $I$ that satifies  $\Sigma$ satisfies $\sigma$ as well. Surely if $\Sigma \models \sigma$ then $\Sigma \models_f \sigma$. \\
\textbf{Fact 1:} Implication and finite implication coincide for FDs. \\
An FD $X \longrightarrow Y$ is said to be \textit{reduced} if there is no
proper subset $W \subset X$ such that $\Sigma \models W \longrightarrow Y$. 
We say $\Sigma$ is \textit{reduced} if every FD in it is reduced. The algorithm  that follows outputs a reduced equivalent to a given set $\Sigma$ of FDs.
\begin{program}
\texttt{Algorithm}\; REDUCED(\Sigma)
\BEGIN
   \Delta \leftarrow \Sigma
   \texttt{for}\, \, (each \,\,  FD \,\, X \longrightarrow Y\, \, in \, \, \Delta ) \; \; \texttt{do}
\hspace{8mm}   \texttt{for}\, \, (each \,\, attribute \,\, A \, \, in \, \,X  ) \; \; \texttt{do}
   \hspace{8mm} \texttt{if} \, \, \Delta \; \models \, X -A \longrightarrow Y
\hspace{8mm} \texttt{then}\, \,  X \leftarrow X-A \,\,in \,\, X \longrightarrow Y    
\texttt{return}\,\, \Delta
\END
\end{program}
Algorithm $REDUCED(\Sigma)$ depends on a test for implication of fds which
determines its complexity.
\subsection{Formal system for functional dependencies}
A formal system for FDs comprises a set of axioms and inference rules -- this was first studied by W.W.Armstrong in 1974 when the significance of implication was not apparent. Armstrong's system denoted by $\textbf{F-A}$ consists of one axiom and three inference rules.
\begin{quote}
\begin{enumerate}
\item[FDA0:] (Reflexivity)  $\quad \vdash  X \longrightarrow X$.
\item[FDA1:] (Transitivity)  $\quad X \longrightarrow Y,  Y \longrightarrow Z \vdash  X \longrightarrow Z$.
\item[FDA2:] (Augmentation and projection)  $\quad X \longrightarrow Y \vdash 
W \longrightarrow Z$ if $X \subseteq W$ and $Z \subseteq Y$. 
\item[FDA4:] (Union) $\quad X \longrightarrow Y, Z \longrightarrow W \vdash
 XZ \longrightarrow YW$.
\end{enumerate}
\end{quote}
\noindent
Describing $\textbf{F-A}$ requires the notion of a derivation --  a derivation of $\sigma$ from $\Sigma$ is denoted by $\Sigma \vdash \sigma$. In a formal system such as $\textbf{F-A}$, given a set $\Sigma$ of FDs  and an FD $\sigma$, by a \textit{derivation} of $\sigma = \sigma_1, \ldots \sigma_n $ we mean: each $\sigma_i$ $(1 \le i \le n)$ is either an instance of an axiom scheme or it follows from the preceding dependencies in the sequence by one of the inference rules.  Soundness and completeness of any system as $\textbf{F-A}$ are expressed as
\begin{quote}
\begin{enumerate}[(1)]
\item  $\textbf{F-A}$ is \textit{sound} if $\Sigma \models \sigma$ is a necessary consequence of $\Sigma \vdash \sigma$. 
\item  $\textbf{F-A}$ is \textit{complete} if $\Sigma \vdash \sigma$ is a necessary  consequence of $\Sigma \models \sigma$.
\end{enumerate} 
\end{quote} 
The formal system \textbf{FD} given below is as in \cite{vard}. 
\textbf{FD} consists of FD1, FD2 and FD3.
\begin{quote}
\begin{enumerate}
\item[FD1:] (Reflexivity)  $\quad \vdash  X \longrightarrow \phi$.
\item[FD2:] (Transitivity)  $\quad X \longrightarrow Y,  Y \longrightarrow Z \vdash
X \longrightarrow Z$.
\item[FD3:] (Augmentation)  $\quad X \longrightarrow Y \vdash XZ \longrightarrow YZ $. 
\end{enumerate}
\end{quote} 
\textbf{Theorem 1:} \textbf{FD} is sound and complete. \\
\textbf{Proof:} Soundness and completeness are proved separately as given below. \\
\textit{Soundness:} As FD1 is vacuously true it is sufficient to show that individually FD2 and FD3 are sound. Let $I$ be a relation on $R$, let $X,Y,Z$ be drawn from $R$   and  $u,v \in I$ be any two tuples. First, considering FD2, assume that $I$ satisfies $X \longrightarrow Y$ and  $Y \longrightarrow Z$. Then (using the definition of FD) if $u[X]=v[X]$ since $u[Y]=v[Y]$ we have $u[Z]=v[Z]$. That is $I$ satisfies
$X \longrightarrow Z$ and hence FD2 is sound. Next, considering FD3, assume that $I$ satisfies $X \longrightarrow Y$. If $u[XZ]=v[XZ]$ then $u[X]=v[X]$ and $u[Z]=v[Z]$. As $I$ satisfies $X \longrightarrow Y$ we have $u[Y]=v[Y]$ and so $u[ZY]=v[ZY]$. In other words $I$ satisfies $XZ \longrightarrow YZ$ and thus FD3 is sound.  \\
\textit{Completeness:} Let $\Sigma$ be the set of FDs that any legal relation $I$ on $R$ needs to satisfy. If $\sigma = X \longrightarrow Y$ be any given FD, completeness requires that if $\Sigma \models \sigma$ then $FD1, FD2$ and $FD3$ are sufficient to conclude $\Sigma \vdash \sigma$. This amounts to proving the contrapositive viz., if $\Sigma \not\vdash \sigma$ then $\Sigma \not\models \sigma$.
In the context $\Sigma$ we define the closure $X^+$ of $X$ as $X^+=\{A \, \vert \, \Sigma \, \vdash \, X \longrightarrow A \}$. By $FD1$, $\vdash  X \longrightarrow \phi$; now invoking $FD3$ taking $Z$ as any $A \in X$ we have 
$\vdash  X \longrightarrow A$. Hence $X \subseteq X^+$. If $X^+=X$ then any relation $I$ that satisfies $\Sigma$ clearly satisfies any given fd $X \longrightarrow Y$. Hence we need to consider the case where $X \subset X^+$. The following  holds due to FD2 and FD3: $\, W \longrightarrow Z_1,  W \longrightarrow Z_2 \vdash W \longrightarrow Z_1Z_2$. Repeated use of this yields $\Sigma \vdash X \longrightarrow X^+$ since $\Sigma \vdash X \longrightarrow A$ by definition, for all $A \in X^+$ and since $X \subseteq X^+$.
If as assumed $\Sigma \not\vdash \sigma$ then we claim that $Y \not \subset X^+$. If possible let the
contrary hold viz., $Y \subset X^+$. In such a case we can use FD1 and FD3 to show that 
$\Sigma \vdash X^+ \longrightarrow Y$. Combining this with the fact $\Sigma \vdash X \longrightarrow X^+$
by virtue of FD2 we get $\Sigma \vdash X \longrightarrow Y$ which is a contradiction. Therefore there
exists some $B \in R$ such that $B \in Y$ but $B \not\in X^+$. We construct a specific legal relation $I$
consisting of two tuples $u,v$. We prescribe that $u[A]=v[A]$ iff $A \in X^+$. In more details
let $u[A]=\alpha$ for all $A \in R$ and $v[A]=\beta$ for any $A \in R \ X^+$. As $X \subset X^+$ 
$u[X]=v[X]$ but by construction $u[Y] \neq v[Y]$. So in $I$ $X \not
\longrightarrow Y$. We now show
that $I$ satisfies $\Sigma$. Let $S \longrightarrow T$ be any FD in $\Sigma$. Indeed if $I$ is legal then
if $u[S]=v[S]$ then we should be able to show $u[T]=v[T]$. So suppose that $u[S]=v[S]$. Then by a previous
argument $S \subset X^+$. Using FD1 and FD3 $X^+ \longrightarrow S$. By the assumption 
$S \longrightarrow T$, with FD1 we can conclude $S \longrightarrow T \vdash X^+ \longrightarrow T$. For any $A \in T$ by FD1 $\, \vdash T \longrightarrow A$.
Then by FD2 it follows that for all $A \in T$ $\; X^+ \longrightarrow A$ which
implies that $\Sigma \vdash X^+ \longrightarrow T$. Therefore $T \subseteq X^+$
and by construction $u[T]=v[T]$. Therefore $I$ satisfies $\Sigma$. Thus $I$ is
one relation that satisfies $\Sigma$ but it does not satisfy $X \longrightarrow Y$ i.e., $\Sigma \not\models \sigma$. \qed
\noindent
\textbf{Remark 3:}\\
i) The counter-example constructed in the proof above is finite. It then follows that in the case of FDs implication and finite implication coincide.   \\
ii) The polynomial-time algorithm for implication of FDs due to C.Beeri and P.A.Bernstein depends on the efficient construction of the closure $X^+$ w.r.t. $\Sigma$ (otherwise denoted as $\Sigma$-Closure$(X)$ or $X^+_{\Sigma}$) also denoted by $cl_{\Sigma}(X)$. \\
iii) Let $\Sigma^*$ denote the {\it semantic closure} of $\Sigma$ i.e.,$\Sigma^* = \{ \sigma \vert \Sigma \models \sigma \}$. Given $\Sigma$ an {\it Armstrong relation} for it is a single relation that satisfies every FD in $\Sigma^*$ and violates every FD not in $\Sigma^*$ (see also \cite{famos} for a generalized definition). Hence an FD $\sigma$ not a member of $\Sigma$ belongs to $\Sigma^*$ iff it is satisfied by an Armstrong relation for $\Sigma$. Armstrong relations represent a case of example-based reasoning an approach known as {\it design-by-example} used in database design (see \cite{maniurai,sven1} for details).  \\
iv) The interaction of FDs in $\Sigma$ is very different when FDs are considered with nulls allowed in relations.  We consider the following example relation which illustrates that under \textit{no information} interpretation of nulls FD2 no longer holds.
\begin{center}
\begin{tabular}{|l|l|r|}
\hline
A      & B     & C  \\ \hline
$a_1$  & $\O$  & $c_1$   \\ 
$a_1$  & $\O$  & $c_2$   \\ \hline
\end{tabular}
\end{center}

\textbf{Lemma 2:} Let $\Sigma$ be a set of FDs to be satisfied by all legal relations over $R$. Let $X,Y \in R$. Then the following holds.\\
\hspace*{5cm} $\Sigma \models X \longrightarrow Y$ $\Longleftrightarrow$
$Y \subseteq cl_{\Sigma}(X)$.   \qed  
\noindent
Thus in order to test if a given FD  $X \longrightarrow Y$ is implied by $\Sigma$ it is sufficient to build $cl_{\Sigma}(X)$ and see if $Y \subseteq cl_{\Sigma}(X)$. We call an FD $X \longrightarrow Y$ \textit{closed} if $Y = cl_{\Sigma}(X)$. A set $\Delta$ of FDs is \textit{closed} if every FD in it is closed. The case $\vert Y \vert=1$ is noteworthy. An fd $X \longrightarrow Y$ is said to be in \textit{canonical form} if $\vert  Y \vert = 1$.  Any FD $X \longrightarrow Y$ can be converted to a set of FDs with each FD in canonical form in view of the following lemma. \\
\textbf{Lemma 3:} For $1 \le j \le k$ and $Y=A_1 \ldots A_k$, $X \longrightarrow Y \models X \longrightarrow A_j$ and $\{ X \longrightarrow A_j \} \models 
X \longrightarrow Y$. 

\subsection{Computing the closure}
The following algorithm $CLOSURE(\Sigma, X)$ takes as input a set of FDs $\Sigma$ and an attribute set $X$ and outputs $cl_{\Sigma}(X)$.
\begin{program}
\texttt{Algorithm}\; CLOSURE(\Sigma, X)
\BEGIN
   Y \leftarrow X  
   \texttt{while}\; \; (there \; exists \; an \; FD \, S \longrightarrow T \, such \; that \; S \subseteq Y \; and \; T \not\subseteq Y ) \; \; \texttt{do}
   \hspace{8mm} Y  \leftarrow  YT     
    \texttt{return}\,\, Y
\END
\end{program}
\noindent
\textbf{Lemma 4:} \texttt{Algorithm}\, $CLOSURE(\Sigma, X)$  correctly outputs
$cl_{\Sigma}(X)$.  \\
\textbf{Proof:} The algorithm terminates after examining a finite number of FDs.
By induction on the steps of the algorithm we first claim that starting from initialization till termination $Y \subseteq cl_{\Sigma}(X)$ holds. As $Y$ is initialized to $X$ the claim is true initially since $\Sigma \models X \longrightarrow A$ for all $A \in X$. Let the claim be true at some intermediate stage during the execution after which the algorithm considers an FD $S \longrightarrow T$ from $\Sigma$ such that  
$S \subseteq Y$. Then $Y \longrightarrow S$ and since $S \longrightarrow T$ we have $Y \longrightarrow T$. By induction hypothesis $X \longrightarrow Y$
and so $X \longrightarrow T$ and we have $X \longrightarrow YT$. That is $\Sigma \models X \longrightarrow YT$. Therefore the claim is true upon termination of the algorithm. \\
We further show that upon termination it is impossible to have $Y \subset cl_{\Sigma}(X)$. If possible let the contrary hold. That is let $B \in cl_{\Sigma}(X)$ but $B \not\in Y$. Since $B \in cl_{\Sigma}(X)$ we  have $\Sigma \models X \longrightarrow B$. We now build a relation $I$ that is legal. Let $I$
consist of two tuples $u,v$ such that $u[A]=v[A]$ iff $A \in Y$. To show that $I$ satisfies $\Sigma$ we assume the contrary if possible. Let $S \longrightarrow T$ be an FD causing violation. Then when this FD was considered by the algorithm we should have had $S \subseteq Y$ but $T \not\subseteq Y$. However in such a 
case the algorithm should have been at some intermediate state of execution. Therefore $I$ satisfies $\Sigma$ but by construction $I$ does not satisfy $X \longrightarrow B$. In symbols $\Sigma \not\models X \longrightarrow B$ which is a contradiction. Hence it follows that $Y = cl_{\Sigma}(X)$.  \qed

\noindent
\textbf{Remark 4:} It follows that $CLOSURE(\Sigma, X)$ can be implemented efficiently. A theorem due to C.Beeri and P.A.Bernstein asserts that the
implication problem for FDs is solvable in time linear in the length of the input. 

\subsection{Covers}
Given two sets of FDs $\Delta$ and $\Sigma$, we say  one is a \textit{cover} for the other if $\Delta \equiv \Sigma$. We can speak of minimal covers in the sense that for any other cover $\Gamma$ of $\Sigma$ a cover $\Delta$ is minimum if 
the number of FDs in $\Delta$ is not greater than that in  $\Gamma$.
With an efficient algorithm for implication we can efficiently determine equivalence, redundancy and a non-redundant cover. Correctness of the following algorithm $NONREDUND(\Sigma)$ is evident.
\begin{program}
\texttt{Algorithm}\; NONREDUND(\Sigma)
\BEGIN
   \Delta \leftarrow \Sigma
   \texttt{for}\, \, (each \,\,  FD \,\, \sigma \, \, in \, \, \Delta ) \; \; \texttt{do}
   \hspace{8mm} \texttt{if} \, \, \Delta - \{\sigma \}\; \models \, \sigma \, \,
\hspace{8mm} \texttt{then} \, \, \Delta = \Delta -  \{\sigma \}     
    \texttt{return}\,\, \Delta
\END
\end{program}
The above algorithm does not necessarily find the minimum covers. 
Let $\Sigma$ be a set of FDs. The following theorem from \cite{sho} (which has a stronger version) is stated without proof.  \\
\textbf{Theorem 2:} Let $\Delta$ be a non-redundant cover for $\Sigma$. If
$\Delta$ is closed then it is minimum.  \qed  \noindent
The following simple but non-trivial algorithm $MINCOVER(\Sigma)$ takes as input a set $\Sigma$ of FDs and outputs a minimum cover for $\Sigma$.
\begin{program}
\texttt{Algorithm}\; MINCOVER(\Sigma)
\BEGIN
   \Delta \leftarrow \Sigma; 
   \texttt{for each} \, \, \sigma \,=\, X \longrightarrow Y \in \Sigma \, \, \texttt{do}
\BEGIN
  \Delta \leftarrow \Delta - \sigma
 \texttt{if} \, \, Y \not \subseteq  CLOSURE(\Delta,X) \,\,\texttt{then}
    \BEGIN     
     Z  \leftarrow CLOSURE(\Delta,Y)
     \Delta \leftarrow \Delta \cup \{X \longrightarrow Z \}
   \END
\END   
    \texttt{return}\,\, \Delta
\END
\end{program}
Algorithm $MINCOVER(\Sigma)$ scans through all the FDs in $\Sigma$. On encountering each
$\sigma$ the algorithm removes $\sigma$ from $\Delta$ and thus updates $\Delta$.
It checks if the removal is safe in which case the updated $\Delta$ will be
equivalent to $\Sigma$. Otherwise it updates $\Delta$ so that after the updation
$\Delta \equiv \Sigma$. The later updation ensures that the
newly added FD is closed. Theorem 2 assures the correctness of $MINCOVER(\Sigma)$. The time complexity of the algorithm can be estimated as
$O(\vert \Sigma \vert \times \tau)$ where $\tau$ is the worst-case time for
one execution of $CLOSURE(\Delta,X)$.\\ 
\textbf{Remark 5:} \\
i) Given a set $\Sigma$ of FDs, by $CANONICAL(\Sigma)$ we denote a \textit{canonical cover} of $\Sigma$ defined as a nonredundant cover for $\Sigma$ such that for any FD $\sigma \in CANONICAL(\Sigma)$ $\sigma$ is reduced and is in canonical form. Canonical covers are not unique \cite{jens}.\\*
ii) Let $U$ denote the set of all possible attributes and let $U \supset R= \{A_1, \ldots A_k \}$ be a relation scheme on which all relations are defined. Let the legal relations be those that precisely satisfy $\Sigma$. Let $\Delta$ be a non-redundant closed cover for $\Sigma$. For $j=1, \ldots ,k$ let $\textbf{K} =\{ X_k \}$ be such that for any $X \in \textbf{K} \,\, \Delta \vdash X \longrightarrow R$ and there is no other $Y$ such that $\Delta \vdash Y \longrightarrow R$. Finding $\textbf{K}$ is $NP-$Complete.
\section{Database schema design}
The principal goal in database schema design is how to design a set of relation schemes to constitute a database and how to indicate their meaningfulness by specifying appropriate constraints such as fds. Intuitively it appears that there is a trade-off between  updating a database versus querying  it -- smaller relational schemes are easier to update while queries on them them may be harder to process. Past research along these lines have focussed on arriving at acceptable ways of grouping attributes into tables and on obtaining normal forms \cite{fa}. The criteria for acceptance of a design is preservation of both information and suggested dependencies and elimination of redundancy.   

In formal terms, a relation scheme is a 2-tuple $(R, \Sigma)$ where $R \subseteq U$ and $\Sigma$ are respectively a relation scheme and an associated set of fds over $R$. Then a database schema $\mathcal{D}$ is a set of relation schemes i.e., $\mathcal{D} \, = \, \{ (R_1, \Sigma_1), \cdots ,(R_k, \Sigma_k) \} $ where $U = \union_{j=1}^{k}$. It is also convenient to define $\Sigma = \union_{j=1}^{k} \Sigma_j$. Finally a database $\mathcal{B}$ over $\mathcal{D}$ is an assignment of a meaningful relation to each relation scheme in each 2-tuple in $\mathcal{D}$. The primary objective in database schema design is that problems such as Codd's anomalies should not exist during database operations w.r.t tuples like additions, deletions and updations -- this concern has resulted in what are called as \textit{normal forms} of database schemas. 
\subsection{Boyce-Codd Normal Form}
For a relation scheme $R$ and attribute set $X \subseteq R$ is a \textit{determinant} of $R$ if there exists at least one attribute $A \in R - X$ such that $\Sigma \models X \longrightarrow A$. If for all  $A \in R$ we have $\Sigma \models X \longrightarrow A$ then $X$ is called a \textit{key} of $R$. If $X$ is a key (also referred to as a superkey) and further if for any $B \in X$, $\Sigma \models X-B \not\longrightarrow A$ then $X$ is called a \textit{minimal key} of $R$. Using an algorithm for implication, the following algorithm constructs minimal keys for a relation scheme $(R,\Sigma)$. 
\begin{program}
\texttt{Algorithm}\; MINIMAL-KEY(R,\Sigma)
\BEGIN
   X \leftarrow R
   \texttt{for}\, \, (each \,\,  A \,\, \in \, \, R ) \; \; \texttt{do}
   \hspace{8mm} \texttt{if} \, \, \Sigma \models X-A \longrightarrow R
\hspace{8mm} \texttt{then} \, \, X = X - A      
    \texttt{return}\,\, X 
\END
\end{program}
It can be observed that with respect to a given relation scheme $(R,\Sigma)$ the set of all minimal keys form a {\it Sperner system}. Let $X = \{x_1, \cdots , x_m \}$ be a ground set with $m \ge 2$ and for $r \ge 2$ let $\mathcal{S} = \{S_1, \cdots ,S_r \} \subset 2^X$. For $1 \le i,j \le m$ if $S_i \subset S_j$ holds for no $i,j \,(i \ne j)$ then $\mathcal{S}$ is referred to as a Sperner system. A theorem due to E.Sperner (1928) asserts that the number of elements in $\mathcal{S}$ is at most $\binom{m}{\lfloor m/2 \rfloor}$. The maximum is attained when $\mathcal{S}$ has all the possible $\lfloor m/2 \rfloor$-element subsets from the ground set.  Further combinatoral aspects and associated probabilistic results concerning data distributions can be found in \cite{demet1}. \\*

In the database process it is useful to know all the minimal keys for a relation scheme $(R,\Sigma)$. The algorithm
ALL-MINIMAL-KEYS is a brute-force approach. \\*

\begin{program}
\texttt{Algorithm}\; ALL-MINIMAL-KEYS(R,\Sigma)
\BEGIN
   \texttt{Let} \, \, P = \mathcal{P}^R
   \WHILE \, \, P \ne \phi \, \, \DO  
   \BEGIN
     \texttt{Let} \, \,  X \in P
     \IF  \Sigma \models X \rightarrow R \, 
     \THEN 
     \BEGIN
          Z = NONREDUND(X)
          \texttt{Output Z as a minimal key}
          \texttt{Let} \, \, S_1, . . . , S_r \in P \,\texttt{such that} \, \, X \subset S_i \, \texttt{for all i=1,...,r}
          P = P-X-S_1- ... -S_r
     \END  
     \ELSE  P = P-X
  \END  
\END
\end{program}

A modification to the algorithm ALL-MINIMAL-KEYS is to combine the test $\Sigma \models X \rightarrow R?$ with the generation of the subsets of $\mathcal{P}^R$ - the basic idea follows. Recall that the subsets of $\mathcal{P}^R$ can be organized as a Hasse diagram in levels of subsets of cardinality 0, 1, . . ., $\vert R \vert$. In an implementation, generate the elements of $\mathcal{P}^R$ levelwise starting with level 1. Then, if a subset $X$ at level $i$  $(1 \le i < \vert R \vert)$ is declared as a key, with respect to the Hasse diagram, further $X$-reachable nodes of type $Y$ such that $X \subset Y$, are not generated. $X$ is enhanced to the $(i+1)$-level subsets if $\Sigma \models X \rightarrow R$ does not hold. \\

We now state the Boyce-Codd normal form ($BCNF$) probably what is referred to as the strongest of all normal forms sought after in database schema design. Assume that the legal relations are associated with sets of fds in canonical form.  \\
\noindent
\textbf{Definition of $BCNF$:} $\mathcal{D}$ satisfies $BCNF$ if whenever $X$ is a determinant of $R$ then $X$ is a key of $R$ where $R$ is a part of any relation scheme of $\mathcal{D}$. 
This implies that for an $A \in R-X$ whenever $X \longrightarrow A$ holds we necessarily have $X$ as a minimal key. \\
Formally, from $\mathcal{D}$ we seek to obtain $\mathcal{E}$ in $BCNF$. For an 
attribute set $X \subset R$ in the context $(R,\Sigma)$ we define the projection of $\Sigma$ on $X$ as $\pi_X(\Sigma)$ = $\{W \longrightarrow A \in \Sigma \, \vert \, \, \, \text{and} \, \,WA \subseteq X  \}$. Let $\Sigma_j=\pi_{R_j}(\Sigma)$. For each $X \longrightarrow A \in \Sigma_j$ if $\Sigma_j \models X \longrightarrow R_j$ then $(R_j,\Sigma_j)$ is in $BCNF$.  

\subsection{Normalization via decomposition}
Let $R=ABCDE$ be a relation scheme such that for any meaningful relation on $R$ the FD $E \longrightarrow CD$ holds. Consider the decomposition of $R$ as $ABE$ and $CDE$.  The following relation $I$ on $R$ is built by ensuring that $E \longrightarrow CD$ holds and by randomly filling values for $A$ and $B$ from their underlying domains.     
\begin{quote}  \begin{quote}  \begin{quote} \begin{quote}  \begin{quote}
{\small
\begin{verbatim}
      I:    -----------------
            A   B   C   D   E
            -----------------
            0   0   0   0   1
            0   1   0   0   1
            1   0   1   1   0
            1   1   1   1   0
            -----------------
\end{verbatim}
}
\end{quote}    \end{quote}   \end{quote}  \end{quote}   \end{quote}   
We can check that 
$\pi_{ABE}(I) * \pi_{CDE}(I)$ = $I$ is always true. This isn't a coincidence  --   the presence of FDs can guarantee non-lossy decompositions as asserted by the following theorem.  \\
\textbf{Theorem 3:} Let $I$ be defined on $R=XYZ$ such that $I$ satisfies $X \longrightarrow Y$. Then the
decomposition of $I$ into $\pi_{XY}(I)$  and $\pi_{XZ}(I)$ is lossless. \\
\textbf{Proof.} Let $J \,= \, \pi_{XY}(I) \Join \pi_{XZ}(I)$. In general $I \subseteq J$. It is therefore sufficient to prove $J \subseteq I$. This is done by showing that $u \in J$ $\Longrightarrow$ $u \in I$ for any $u \in J$. By the definition of $J$, $u[XY] \in \pi_{XY}(I)$ and $u[XZ] \in \pi_{XZ}(I)$. As
$\pi_{XY}(I)$ and $\pi_{XZ}(I)$ are obtained from $I$, there are two tuples $v,w \in I$ such that $u[XY]=v[XY]$ and $u[XZ]=w[XZ]$. Consequently $v[X]=w[X]$ and since $X \longrightarrow Y$ it follows that $w[XY]=v[XY]=u[XY]$. We reason $w[XYZ]=u[XYZ]$ i.e., $w=u$.  \qed 

To implement normalization via decomposition assume that $\mathcal{D}$ is not in $BCNF$ where $\mathcal{D}$ is a database schema defined as above.. W.l.o.g. assume that the relation scheme $(R_j, \Sigma_j)$ is the cause for violation. 
Since $(R_j, \Sigma_j)$ is not in $BCNF$ there exists an $X$ that is a determinant of $R_j$ but that isn't a key for $R_j$. Therefore there is an attribute $A \in R_j-X$ such that $\Sigma \models X \longrightarrow A$ where
$\Sigma = \union_{j=1}^{k} \Sigma_j$. In the decomposition process we invoke theorem 3  and replace $(R_j, \Sigma_j)$ by $D_1=(\pi_{XA}(R_j), \Sigma_j^1)$ and $D_2=(\pi_{R_j-A}(R_j),\Sigma_j^2)$ where $\Sigma_j^1$=$\pi_{XA}(\Sigma_j)$ and $\Sigma_j^2$=$\pi_{R_j-A}(\Sigma_j)$ and continue further the process of decomposition if $D_2$ is not in $BCNF$.

Let $\mathcal{D}$ be as defined above and let $\mathcal{E}=(U,\Sigma)$, also referred to as a universal schema. As a part of the design process we would like to find when does it make a reasonable sense to say that $\mathcal{D}$ represents $\mathcal{E}$. It perhaps follows intuitively that we need these two conditions viz., (a) there should be no loss of information if relations are stored using schema $\mathcal{D}$ rather than as schema $\mathcal{E}$ and (b) all the  $\Sigma$ should logically imply the fds in all the $\Sigma_i's$ and together the $\Sigma_i's$ should logically imply $\Sigma$. This amounts to the requirement that a decomposition from $\mathcal{E}$ to $\mathcal{D}$ should
be \textit{lossless} and \textit{dependency preserving}. Formally $\mathcal{D}$ \textit{represents} $\mathcal{E}$ if the following  conditions hold together.
\begin{enumerate}[(i)]
\item Let $I$ be a relation on $U$ satisfying $\Sigma$. Then for $j=1, \ldots k$ the decomposition of $I$ into $\pi_{R_j}(I)'s$ is lossless.
\item Let $\Delta = \union_{j=1}^{k} \Sigma_j$. Then $\Sigma \models \Delta$ and $\Delta \models \Sigma$. 
\end{enumerate} 
The decomposition in theorem 3 is such that (i) is guaranteed but only half of (ii) is satisfied. Unfortunately in general it appears that it not possible to efficiently find a decomposition resulting in $BCNF$ that satisfies both (i) and (ii) above. This is asserted by theorem 4 that follows.  

\subsubsection{Checking for $BCNF$ violations}
We begin with the hitting set problem which is $NP-$complete
(problem [SP8] on p.222 of \cite{garey}).\\
\textsc{Hitting set:}
Let $T=\{ A_1, \ldots ,A_n \}$ and let $B_j \subset T$ for $j= 1, \ldots ,m$. The problem asks to find
if possible a \emph{hitting set} $W \subseteq T$ such that for each $j$, $\vert W \cap B_j \vert = \alpha$ where $\alpha \ge 1$.  \\
In the following example $\alpha = 1$. \\ 
Ex. Setting $T = \{ p_1, \ldots ,p_8 \}$, let $B_1= \{p_1,p_2,p_3 \}$,  $B_2= \{p_2,p_3,p_4 \}$, $B_3= \{p_1,p_7,p_8 \}$,  $B_4= \{p_5,p_6,p_7 \}$. We interpret $T$ to be a set of persons and we set that a task $t_j$ requires for its completion skills available with any person in $B_j$. It is required to
find a group of persons from $T$ who can complete all the tasks subject to the constraint that only one person is selected from each $B_j$. We may require finding 
(i)  a $W$ so as to maximize or minimize $\vert W \vert$ or 
(ii) $W_1$ and $W_2$ where possible so that $W_1 \cap W_2 = \phi$.\\
The problem of determining if a given database schema violates $BCNF$ is a hard 
problem. The following proof relies on the hardness of \textsc{Hitting set} with $\alpha = 1$.  \\

\textbf{Theorem 4:} [C.Beeri and P.A.Bernstein]  Let $\mathcal{D}$ be a given database schema. It is $NP$-complete to check if there is a $BCNF$ violation in
$\mathcal{D}$.\\
\textbf{Proof:} That the problem belongs to the class $NP$ is clear from
\cite{vard}. Following \cite{vard} we show that the problem is $NP-$hard. That is, we reduce the hitting set problem to the problem at hand. More specifically, the proof shows that each instance of the hitting set problem can be mapped in polynomial-time to a database schema such that there exists a hitting set iff the produced schema violates $BCNF$. 

Let $U=\{ A_1, \ldots , A_n,B_1, \ldots, B_m, C, D \}$ where we retain the notations of the hitting set problem. We build a database schema $\mathcal{D}$ consisting of the following relation schemes. \\*
D1. For every pair $A_i,B_j$ such that $A_i \in B_j$ we include in $\mathcal{D}$ $\{A_iB_j, (A_i \longrightarrow B_j) \}$. \\*  
D2. We include in $\mathcal{D}$ $\{ B_1 \ldots B_m CD , (B_1 \ldots B_m \longrightarrow C) \}$.\\*  
D3. Finally we include in $\mathcal{D}$ $\{ A_1 \ldots A_n CD, ( \{ CD \longrightarrow  A_1 \ldots A_n \}$
$\cup  \{ A_iA_j \longrightarrow CD \,\, \text{if}\, \, i \ne j$ and both  $A_i,A_j \, \,\text{belong to}\, \, B_k \, \, \text{for some}\, \,  k  \}  ) \}$.\\*
It is not difficult to reason that with D1, D2 and D3, $\mathcal{D}$ can be constructed in polynomial-time. Let $W = \{ A_{\alpha_1}, \ldots , A_{\alpha_r} \} \, \subseteq \, T$ be a hitting set. Then for every $B_j$  $W \cap B_j = A_{\alpha_k}, \, \, 1 \le k \le r$. By D1 this necessarily means $A_{\alpha_k} \longrightarrow B_j$. We can conclude, for every $i$, $\Sigma \models W \longrightarrow B_i$. Using lemma 4 with D2 we then have $\Sigma \models W \longrightarrow C$.
So in D3 $W$ is a determinant of $A_1 \ldots A_nCD$. We now make the following claim 4.1 which implies that $W$ is not a key for the relational scheme in D3.\\
Claim 4.1: $cl_{\Sigma}(W) = WB_1 \ldots B_mC$ where $\Sigma$ is the set of all fds in $\mathcal{D}$. \\
To establish the claim we note that from $CLOSURE(\Sigma,X)$ it follows that it is sufficient to show that for every fd
$S \longrightarrow T \in \Sigma$ either $S \not \subseteq cl_{\Sigma}(W)$ or $T \subseteq cl_{\Sigma}(W)$
holds. This follows by considering D1, D2 and D3 separately.  

Conversely let $\mathcal{D}$ be not in $BCNF$. Then one or more of $D1, D2, D3$
should contain a violation of $BCNF$ -- we find that only $D3$ can cause a violation. From $D3$ we first observe that for any $i,j \,\,(i \ne j)$ if $A_i
\in B_k$ and $A_j \in B_k$ then $A_iA_j \longrightarrow A_1 \ldots A_n$. Let
$W \subseteq A_1 \ldots A_nCD$ be a determinant but not a key.  Clearly
$C,D \not\in W$. So let $W \subseteq A_1 \ldots A_n$. Because of the observation above $W$ cannot contain two distinct $A_i, A_j$ belonging to some $B_k$. That
is $W$ consists of $A_i$'s such that every such $A_i \in B_j$ and there exists
no other $A_j \in B_j$. Let $W = \{ A_{\alpha_1}, \ldots , A_{\alpha_r} \}$ so that each $A_{\alpha_j}$ embraces some $B_p$. If all the $B_i$'s are not embraced by our choice of $W$ let there be a $B_l$ such that there exists $A_q \in B_l$  as $B_l \subset T$ but $A_q \not\in W$. We then update the current $W$ by including $A_q$ and do not further reckon any $B_s$ if $A_q \in B_s$. This way we can expand $W$ embracing all possibly  left out $B_i$'s so that it becomes a hitting set.  \qed

\subsection{3NF and normalization via synthesis}
We assume $\mathcal{D}$ is a database schema as defined above.
Starting from  $\mathcal{D}$ instead of obtaining a desirable database schema in $BCNF$ representing $\mathcal{D}$ which may not be feasible in some cases it is possible to settle for a weaker normal form referred to as the
third normal form $(3NF)$. The definition of $3NF$ can be given in terms of a \textit{strong determinant}. An attribute $A \in R_j$ is called \textit{prime}
if there is a minimal key $Z$ of $R_j$ such that $A \in Z$. $Z$ is a strong determinant of $R_j$ if $Z \subseteq R_j$ and there exists a nonprime $A \in R_j-Z$ such that $\Sigma \models Z \longrightarrow A$. Here it is not necessary for $Z$ to be a key. The following example illustrates the back propagation of primality in a chain of simple FDs.\\
Ex. Let $(R,\Sigma)$ be a relation schema. For $1 \le j \le n$ let $A_j \longrightarrow A_{j+1} \in \Sigma$.  If $A_{n+1} \in K$ for a key $K$ of $R$ then $A_1 \in K$. It is sufficient to show that for an FD $A \longrightarrow B$ in the chain $A$ is prime if $B$ is so. W.l.o.g. assume $B \not\longrightarrow A$. Primality of $B$ means $K-B \not\longrightarrow R$ for a key $K$ such that $B \in K$. If $A \in K$ the result follows. So let $A \not\in K$. Then $\{K \longrightarrow A,B \not\longrightarrow A \} \models K-B \longrightarrow A$ -- a contradiction. \\*

\textbf{Modern definition of $3NF$:} $\mathcal{D}$ is in $3NF$ if whenever $Z$ is a strong determinant for any $R$ in $\mathcal{D}$ then $Z$ is a key of
$R$.\\ 
The problem of determining whether a given relation scheme $(R_j, \Sigma_j)$ is in $3NF$ can be done in one of the following two ways.\\
W1. Show that for all FD $X \longrightarrow A \in \Delta_j$, $X$ is a key or $A$ is prime, where $\Delta_j = CANONICAL(\Sigma_j)$. Conclude that $3NF$ is not violated.  \\* 
W2. Show that there exists an FD $X \longrightarrow A \in \Delta_j$ such that
$X$ is not a key and $A$ is not prime, where $\Delta_j$ is as in W1. 
Conclude that $3NF$ is violated.  \\*
\textbf{Remark 6:}\\ 
i) First we note that $BCNF$ implies $3NF$. Let $(R_j, \Sigma_j)$ be in $\mathcal{D}$ and let $X \longrightarrow A \in \Sigma_j$.  If $A \in R_j$ is prime then there is at least one  minimal key $Y \subseteq R_j$ such that $A \in Y$. Assume that we decompose $R_j$ as $\pi_{XA}(R_j)$ and $\pi_{R_j-A}(R_j)$. Then the fd  $Y \longrightarrow R_j$ is lost since $Y$ is a minimal key and $A \in Y$. Therefore the resulting schema  does not represent $\mathcal{D}$. Such a problem will not arise if every $A \in R_j$ is nonprime.\\*
ii) The problem of determining whether a given attribute is prime is $NP-$Complete. (problem [SR28] on p.232 of \cite{garey})\\*
iii) \cite{worland} presents an algorithm to check whether or not a
relation scheme is in $3NF$. \\*
The following theorem stated without proof gives a condition for a $3NF$ relation to be in $BCNF$. \\*
\textbf{Theorem 5:} Let $(R,\Sigma)$ be in $3NF$. For every pair of minimal keys  $K_1,K_2$ if $K_1 \cap K_2=\phi$ then $(R,\Sigma)$ is in $BCNF$.\\*
To prove the theorem we take $(R,\Sigma)$ in $3NF$ and w.l.o.g. assume that $\Sigma$ is canonical. By W1 any FD $X \longrightarrow A$ should be such that $X$ is a key or $A$ is prime. Letting $R=K_1 \ldots K_nS \;(n > 2)$ where $K_j $ $(1 \le j \le n)$ are the only minimal keys, by a case analysis the result follows. Converse of theorem 5 is not true always. 

Normalization to $3NF$ via decomposition  is not practical but fact 2 is more positive. \\*
\textbf{Fact 2:} For any universal schema $\mathcal{E}$ there exists a 
database schema $\mathcal{D}$ representing $\mathcal{E}$. Moreover normalization through synthesis finds $\mathcal{D}$ efficiently as shown by P.A.Bernstein around 1976.\\*
The $3NF$ synthesis algorithm can be described along the following lines.
 \begin{program}
 \texttt{Algorithm}\; 3NF(U,\Sigma)
 \BEGIN
    \mathcal{D} \leftarrow \phi
   \Delta \leftarrow CANONICAL(U,\Sigma) 
   X \leftarrow MINIMAL-KEY(U,\Delta)
   \texttt{for}\,\,(each \,\, fd \,\,Y \longleftarrow A \,\in \Delta ) \;\; \texttt{do}
   \hspace{8mm} \mathcal{D} \leftarrow \mathcal{D} \cup (YA, \pi_{YA}(\Delta) ) 
   \mathcal{D} \leftarrow \mathcal{D} \union (X, \phi) 
 \END
\end{program}
From the input $\Sigma$ algorithm $3NF(U, \Sigma)$ first finds a canonical cover for $\Sigma$. Then for every FD it a relation scheme is created. Finally a key for $U$ is added. It can be seen that the algorithm is efficient. The following theorem is stated without proof.\\*
\textbf{Theorem 6:} Algorithm $3NF(U,\Sigma)$ is correct: the output $\mathcal{D}$ of the algorithm represents the input schema $(U,\Sigma)$. \\*

\subsection{More design aspects}
Practical database design has also prompted the study of sets of FDs. Let $(R,\Sigma)$ be a given starting point in the design process.
Consider $\Delta \in \mathcal{P}^\Sigma$. One problem is to derive a $\Delta$ such that $\Delta \equiv \Sigma$ and $\vert \Delta \vert$ is the maximum. To add some elaboration we refer to \cite{demetr}. For attributes $A$ and $B$ in $R$ where $A \subset B$ if $A \longrightarrow C \in \Delta$ then  $B \longrightarrow C \notin \Delta$ as $A \subset B$ implies $B \longrightarrow A$ and hence by transitivity $\{ A \longrightarrow C, B \longrightarrow A \} \vdash A \longrightarrow C$. Further reasoning suggests constructive initialization of $\Delta$ with the maximum number FDs from $\Sigma$ of the form $X \longrightarrow Y$ where $Y \subseteq X$ is false. If $R$ is of arity $n$ we then get by Sperner's theorem a theoretical lower bound of $\binom{n}{\lfloor \frac{n}{2} \rfloor}$.  \\*

We consider the following relational scheme for a publishing house. 
\begin{center}
$BOOKS(Bid, Author, Book Title, SizeCode, No.pp)$.  
\end{center}
A relational instance is given in the table below. \\

\begin{center}
\begin{tabular}{|l|l|l|l|r|}
\hline
$Bid$  & $Author$ &      $Book Title$         & $SizeCode$ & $No.pp$   \\ \hline
$111$  & Alice    & Structural Architecture   & Poster-size      & 100 \\ 
$111$  & Alice    & Structural Architecture   & Life-size        & 250 \\ 
$222$  & Carol    & Computer Architecture     & Life-size        & 200 \\ 
$222$  & Carol    & Computer Architecture     & PBack-size       & 600 \\ 
$222$  & Carol    & Computer Architecture     & Pocket-size      & 850 \\ \hline
\end{tabular}
\end{center}

A careful interpretation of the intended real-world meanings of $BOOKS$ suggest the following: the FD $Bid \longrightarrow Author$ should hold while the non-FDs $Bid -/-> Size-Code$ and $Bid -/->No-pp$ should also hold in any relation - here $X -/-> Y$ is a non-FD denoting that the FD $X \longrightarrow Y$ should be excluded. In this context, a relation scheme is specified by the $3$-tuple $(R, \Sigma_1, \Sigma_0)$ where $\Sigma_1$ and $\Sigma_0$ are given below (see \cite{demetrov} for details): 

\begin{description}
\item[$\Sigma_1$] -- the set $\{ \sigma^1_i \vert i= 1, 2, . . . \}$ of valid FDs i.e., $\Sigma_1 \cup \Sigma_0 \models \sigma^1_i$ for all admissible $i$. 

\item[$\Sigma_0$] -- the set $\{ \sigma^0_i \vert i= 1, 2, . . . \}$ of excluded FDs i.e., $\Sigma_1 \cup \Sigma_0 \models \sigma^0_i$ for all admissible $i$.
\end{description}

\noindent
An important design aspect (see \cite{demetrov} for details) is then to start with an initial pair 
$\Sigma_1, \Sigma_0$ and generate the logical closures of FDs and non-FDs in $\Sigma_1, \Sigma_0$ guided by a set of sound and complete axioms. \\

When FDs with nulls are allowed the database design process involves new concerns. We follow the example given in \cite{sven1} illustrating the use of Armstrong relations in understanding the semantics of relational schemes. We consider a school where as part of registration the table $MCS(Sid, Tid, MCrs)$ captures mentors, their students and the only course to be offered by each mentor. To begin with assume that the FDs $Sid \longrightarrow Tid$  and $Tid \longrightarrow MCrs$ are declared to be in $\Sigma$. Further assume that $Sid$ and $Tid$ are declared as {\it not null}. For these constraints the following is an Armstrong relation for $\Sigma$.
\begin{center}
\begin{tabular}{|l|l|r|}
\hline
Sid    & Tid    & MCrs  \\ \hline
$S_1$  & $T_1$  & $C_1$   \\ 
$S_2$  & $T_1$  & $C_2$   \\ 
$S_3$  & $T_2$  & $C_2$   \\ 
$S_4$  & $\O$   & $C_3$   \\ 
$S_4$  & $\O$   & $C_4$   \\ \hline
\end{tabular}
\end{center}

Inspecting the table $MCS$ reveals that neither the FD $Sid \longrightarrow MCrs$ nor the minimal key $Sid$ have been captured. At the design level we reason that the $MCS$ schema is to be refined by adding the FD $Sid \longrightarrow MCrs$ to $\Sigma$ or by specifying the constraint that $Tid$ is not null.  \\*

We refer to \cite{demet} for some modern approaches concerning the presentation of sets of FDs based on which full knowledge on the validity of FDs with respect to a stated context may be extracted. Database theoreticians have also underlined the relevance and importance of horizontal decompositions of structured data and their retrieval, requiring orthogonal considerations in normal form designs arising out of certain applications and/or requirements.    \\*

Most interesting database design problems appear to be intractable as viewed from the classical 
perspective. It seems natural to reason whether there are contexts where the FDs occur in some  restricted sense so as to admit robust fixed-parameter algorithms for some of the database design problems. This practically important aspect is analyzed in \cite{lob}.
   
\section{Some concluding remarks}   
It is known that redundancy and potential inconsistency can also be present in certain relations in the absence of FDs. On a relation scheme $R$ a  \textit{multivalued dependency} ($mvd$) occurs when the values on $X$ determines the set of values on $Y$ independent of the set of values on $R-Y$. Formally if $XY \subseteq R$, $X$ \textit{multidetermines} $Y$ i.e., $X \rightarrow \rightarrow Y$ if for every relation $I$ on $R$, for all tuples $u,v \in I$ if $u[X]=v[X]$ then there exists a tuple $w \in I$ such that $w[X]=u[X]=v[X]$, $w[Y]=u[Y]$, $w[R-XY]=v[R-XY]$. Relations on $R$ satisfy $X \rightarrow \rightarrow Y$ when $R$ is decomposable to its projections on $XY$ and $X(R-Y)$ without loss of information; in other words $R = \pi_{XY}(R) \Join \pi_{X(R-Y)}(R)$.  The fourth normal from ({\it 4NF}), a generalization of $BCNF$ requires that every mvd is a consequence of minimal keys. Let $R$ be a relation scheme and $X,Y \subseteq R$ $(X \ne \phi, Y \ne \phi)$ and let $\Sigma$ be a set of fds and mvds that need to satisfied on legal relations on $R$. $R$ is in {\it 4NF} if for every mvd $X \rightarrow \rightarrow Y$ that is to hold on legal relations over $R$ either the mvd is trivial i.e., $Y \subseteq X$ or $XY=R$ or $X$ is a key in the sense defined before. A theorem states that if $R$ obeys only those $FD$s and mvds that are logical consequences of a set of FDs then {\it 4NF} coincides with $BCNF$. The interaction between FDs and mvds has been studied under a sound and complete formal system.
It is believed that reasoning about a constraint set with different classes and the associated implication problems are significantly harder compared to dealing with the classes separately, possibly due to interactions among the classes.
Similar to theorem 5 another theorem asserts that given $(R,\Sigma)$ in $BCNF$, if there is a simple key then $(R,\Sigma)$ is in {\it 4NF}. See also \cite{sven}. 

{Inclusion dependency} or $\textsc{ind}$ for short, is an integrity constraint of practical relevance that arises in the context of Codd's referential integrity. Attribute sets $X,Y$ of same cardinality $p \ge 1$ are called {\it compatible} if for a given order of attributes of $X$ there is an ordering of attributes of $Y$ so that $Dom(X[j])=Dom(Y[j])$, for $j=1, \cdots ,p$. W.r.t. relation schemes $R_i,R_j$ an $\textsc{ind}$ is a syntactic statement $R_i[X] \subseteq R_j[Y]$ where $X \subset R_i$,$Y \subset R_j$ and $X,Y$ are compatible; if $\vert X \vert = k$ we have a case of {\it unary} $\textsc{ind}$. Let $R_i$ and $R_j$ be a part of a database schema. A $\textsc{ind}$ stated as above is satisfied if for all associated relations $I_i$ and $I_j$ we have: $\forall u \in I_i \, \exists v \in I_j : u[X]=v[Y]$ i.e., $\pi_X(I_i)=\pi_Y(I_j)$. Equivalently (see \cite{fab}) we have $\vert \pi_X(I_i) \vert = \vert \pi_X(I_i) \Join \pi_Y(I_j) \vert = \vert \pi_{X,Y}(I_i \Join_{X=Y} I_j ) \vert.$ With the context described earlier, in an \textsc{ind} specification $R_i[X] \subseteq R_j[Y]$ if $Y$ is a key then $X$ is termed a {\it (minimal) foreign key} -- this notion is useful in natural specifications of a database instances avoiding imprecise representations. We note that $\textsc{ind}s$ admit a complete axiomatization. If $\Sigma$ is a set of $\textsc{ind}s$
and $\sigma$ is a given $\textsc{ind}$ the inference problem  $\Sigma \models \sigma$? is $PSPACE-$Complete. Hence there is no polynomial-time algorithm expected for this inference problem, unless $P=PSPACE$, as shown in \cite{mcas}.         

\end{document}